\documentstyle[aps]{revtex}
\begin{document}
\draft
% \preprint{ }
\title{A geometric, dynamical approach to thermodynamics}
\author{Hans Henrik Rugh }
\address {Department of Mathematics, University of Warwick,
          Coventry, CV4 7AL, England}
\date{\today}
% \twocolumn[
\maketitle
%\widetext
 \begin{abstract}
We present a geometric and dynamical approach to the micro-canonical
ensemble of classical Hamiltonian systems. We generalize the arguments 
in \cite{Rugh} and show that the energy-derivative of a micro-canonical
average is itself  micro-canonically observable. In particular,
temperature, specific heat and higher order derivatives of the entropy
can be observed dynamically. We give perturbative, asymptotic formulas
by which the canonical ensemble itself can be reconstructed from
micro-canonical measurements only. In a purely micro-canonical approach
we rederive formulas by Lebowitz {\it et al} \cite{LPV}, relating 
e.g.\ specific heat to fluctuations in the kinetic energy. We show that
under natural assumptions on the fluctuations in the kinetic energy
the micro-canonical temperature is asymptotically equivalent to the
standard canonical definition using the kinetic energy.
 \end{abstract}
\pacs{05.20.Gg, 05.20.-y, 05.45.+b, 05.70.-a, 02.40.Vh, 02.40.-k}
% ] \narrowtext

\section{Introduction}
For an isolated  classical Hamiltonian system the ergodic hypothesis 
asserts that the time average of an observable 
 along almost any trajectory 
may be replaced by a space average over a suitable 
subset of the phase space, typically the energy surface.
Such an average  is denoted the micro-canonical ensemble average
or for short, $\mu$-average.
The thermodynamic variables in this ensemble 
are the first integrals as well as extensive quantities like volume
and particle numbers
(cf e.g. Abraham and Marsden \cite[Chapter 3.7]{Abraham}).
In the so-called thermodynamic limit 
of many systems coupled weakly 
one expects in equilibrium each individual system  
to behave according to the canonical or Gibbs ensemble.
In this ensemble the 
free parameters \cite{LL} are the
variables conjugated to (some of) the first integrals.
Quite simple statistical principles argue in favor of such
an approach but rigorous results are sparse \cite{Ruelle}.
Whereas the theory of the canonical ensemble has been elaborated 
to almost perfection,  making it a corner stone in modern physics,
our present understanding of the $\mu$-ensemble
and the equivalence of the two ensembles is remarkably incomplete. 
On the other hand modern computer technology makes it ever more
important to understand and give precise interpretations for 
dynamical measurements in the $\mu$-ensemble \cite{EM}.

\mbox{}From a geometrical point of view the $\mu$-ensemble 
is an average of smooth quantities over a (genericly smooth) sub-manifold
of phase space, fixed by the first integrals.
For the sake of clarity we restrict ourselves
to the case where the energy is the only extensive parameter.
But even in this simple picture it is not a priori clear how
to describe the important role played by the variable conjugate
to the energy, i.e.\ the inverse temperature.
When the Hamiltonian contains one or more separate terms of the form
momentum squared 
the canonical average of such a term 
yields precisely the (canonical) temperature.
The time-average of such a term is therefore often used
 as a measure of the physical temperature of the system
(cf. \cite[Example 3.7.27]{Abraham} or \cite{LPV}).
This approach, however, presumes both the ergodic hypothesis and
the equivalence of ensembles.
In \cite{Rugh} we used simple geometrical arguments
to show that the $\mu$-temperature
is in fact measurable in the $\mu$-ensemble itself.
In particular, assuming ergodicity only, we 
constructed explicitly an observable whose
average yields the $\mu$-temperature.
In the present article we shall show that these arguments 
in fact carries over to
a much wider range of $\mu$-observables.

Below we establish (Theorem 1) two fundamental identities 
that will allow us to measure any energy-derivative of a $\mu$-average
{\em within} the $\mu$-ensemble itself.
Thus not only the temperature but also the specific heat and any
higher order derivatives of the $\mu$-canonical entropy can be observed
dynamically. 

We give two main examples to illustrate 
these principles.
In the first we use a perturbative expansion to
show that by measuring all energy derivatives 
in the micro-canonical ensemble we can in principle
reconstruct the canonical ensemble,
cf.\ equations (\ref{eq:pert}) and (\ref{eq:F}).
This approach is, however, based upon a Gaussian expansion 
which itself relies on analyticity. 
On general grounds one would expect
such results to be at best
asymptotic, e.g. in the number of degrees of freedom.

In the second example we consider the thermodynamics
of particles in a box. Lebowitz {\it et al} \cite{LPV}
established relations between 
e.g.\ the specific heat and fluctuations
in the kinetic energy in the micro-canonical ensemble.
We rederive these relations in a
purely micro-canonical approach. We also show that when both the
kinetic energy and the square of
fluctuations in the kinetic energy are of order $N$ our
micro-canonical derivation of temperature is 
asymptotically equivalent 
to the standard canonical definition of temperature
which uses the average kinetic energy. It is of interest to study
deviations in the two approaches for systems out of equilibrium.

\section{Energy Derivatives}
The phase space is
an $n=2N$-dimensional symplectic 
manifold $({\cal M},\omega)$ where $\omega$
is a non-degenerate symplectic two-form and the Liouville
volume form $m = \wedge^N \omega$ is nowhere vanishing.
The standard example is (a subset of) Euclidean space
         ${\cal M} = R^{2N}$ and $m=$ Lebesgue
         measure (cf. Example B below).
A Hamiltonian function $H : {\cal M} \rightarrow R$
 generates \cite{Abraham,Arnold} then
 a vector field $IdH$ and a flow $g^t_{IdH}$ which
preserves the Liouville measure as well as the energy surface,
$\Sigma_E = \{\xi \in {\cal M} : H(\xi) = E \}$.
It follows (\cite[chapter 3.7]{Abraham}) that the flow also
preserves the restricted Liouville measure, formally given by~:

      \begin{equation}
          \mu_E = m \; \delta(H - E) . \end{equation}

The $\mu$-measure of an observable $\phi$ 
is given by
      $ \mu_E (\phi) = \int m \; \delta(H - E) \; \phi$ 
and is a function of the energy.  The
$\mu$-ensemble is the corresponding 
probability distribution, yielding the $\mu$-average of $\phi$
over the energy surface~:
      \begin{equation}
           \langle \phi ; E \rangle = \mu_E(\phi)/\mu_E(1) .\end{equation}
If the Hamiltonian flow is ergodic with respect to
the restricted Liouville measure then 
by Birkhoff's theorem this $\mu$-average equals
the time average of $\phi$ for almost any initial point on
the energy surface $\Sigma_E$. 
We say that $\langle \phi; E\rangle$ is measurable
in the $\mu$-ensemble.

Of particular importance is the $\mu$-entropy
and the associated (inverse) temperature~:
    \begin{equation}
          S(E) = \log \mu_E (1)  \ , \ \  \ 
          \frac{1}{T(E)} = \frac{\partial S(E)}{\partial E}
                 = 
    \frac{\frac{\partial}{\partial E}\mu_E(1)} {\mu_E(1)}
     \label{eq:temp} .\end{equation}
The energy derivative and the $\mu$-averages can now be related
through the following~:\\

\noindent \underline{Theorem :}
\begin{itemize}
\item[ ]
{\em 
Assume that
$\Sigma_E$ is a regular energy surface of the Hamiltonian function $H$
and that $X$ is a vector field defined in a neighborhood of $\Sigma_E$ 
satisfying~:
       \begin{equation}
            dH(X) \equiv 1 .  \label{eq:one}\end{equation}
Then the $\mu$-measure and the $\mu$-average of a observable
$\phi$ verify the identities~:
       \begin{equation}
             \frac{\partial}{\partial E} \mu_E (\phi) =
               \mu_E (\mbox{\rm div}\;(\phi X)) \label{eq:deriv}  , \end{equation}
       \begin{equation}
           \frac{\partial}{\partial E} \langle \phi; E\rangle 
          =  \langle \mbox{\rm div}\;( \phi X) ; E\rangle - 
         \frac{\langle \phi; E\rangle}{T(E)} . 
                 \label{eq:av}\end{equation}
\mbox{ }\\
}
\end{itemize}

\noindent In coordinates, $X\cdot \nabla = X_i \frac{\partial}{\partial x_i}$,
$m=\rho \; dx_1 \wedge \ldots \wedge dx_n$, one has the explicit formulae~:
      \begin{equation}
      \mbox{div} \;(\phi X) = 
     \frac{1}{\rho} \sum_i \frac{\partial}{\partial x_i}
                       (\rho \phi X_i) \ ,
     \ \ \ dH(X) = \sum_i X_i \frac{\partial H}
                                   {\partial x_i} \equiv 1 .
     \label{eq:diver} \end{equation}

\noindent It is sufficient to verify the identity (\ref{eq:deriv})
since the other follows from (\ref{eq:temp}) and
\begin{equation}
    \frac{\partial}{\partial E} \langle \phi; E\rangle =
  \frac{\frac{\partial}{\partial E} \mu_E(\phi)}
              {\mu_E(1)} -
    \frac{\mu_E(\phi)}{\mu_E(1)} \; \frac{\frac{\partial}{\partial E}\mu_E(1)}
      {\mu_E(1)} . \end{equation}

According to Khinchin \cite{Khinchin}, the  $\mu$-canonical
measure is proportional to  $d\Sigma/\| \nabla H \|$ where $d\Sigma$
is the area element on the energy surface. Thus for an observable $\phi$
we have

\begin{equation}
    \int_{H \leq E} m \phi = \int^E_{-\infty} du \int_{H=u} 
         \frac{d\Sigma}{\| \nabla H \|} \phi.
        \label{eq:Khinchin}
\end{equation}
In the language of differential geometry the $\mu$-canonical
measure can be expressed as a 
differential $(2n-1)$-form~:
\begin{equation} 
   \mu = i_X m,
\end{equation}
where $X$ is any vector field for which $dH(X) \equiv 1$.
Provided $H$ has no critical points on the energy surface one
can always find such a vector field in a neighborhood of that
energy surface. Although $\mu$ is not unique, its restriction to
an energy surface is unique and equivalent to the weighted area element in
Khinchin's formula.
The (exterior) derivative of $\mu \phi$ ($\phi$ being an observable)
is given by~:
  \begin{equation}
      d(\mu \phi) = d(i_X m \phi) =
                    d(i_{\phi X} m) = m \; 
                   \mbox{div} (\phi X) , \end{equation}
where the divergence of $X\phi$ was given by equation (\ref{eq:diver})
 above.
Stokes Theorem and the relation $m = dH \wedge \mu$ yields then
      \begin{equation}
         \int_{H=E} {\mu \phi} = \int_{H \leq E} d(\mu \phi)
         = \int_{-\infty}^E du \int_{H=u} \mu \; \mbox{div} (X \phi).
      \end{equation}
Hence, taking a further energy derivative~:
      \begin{equation}
         \frac{\partial}{\partial E} \mu_E(\phi)
                     = \mu_E (\mbox{div} (\phi X)) ,
       \end{equation}
as we wanted to show.

\section{General Remarks}
 
\begin{itemize}
\item Non-uniqueness:
  Note that the vector field $X$ is far from unique.
      One may add any
      vector field tangent to the energy surfaces.
      This corresponds to a reparametrization of the surfaces
      and does not change
      the average of $\mbox{div}\; (\phi X)$. It does, however, affect
      fluctuations in this observable and hence a wise choice of $X$
      could lead to better convergence in numerical experiments.
\item 
      Setting $\phi\equiv 1$ it follows from (\ref{eq:av}) and
      $\langle \phi;E \rangle \equiv 1$ that~:
          \begin{equation}
             \frac{1}{T(E)} = \langle \mbox{div} \; X ; E \rangle .
          \end{equation}
      Neither this formula nor those in the Theorem 
      make any reference to
      metric properties of ${\cal M}$.
      A metric on $M$ allows one to compute gradients of 
      functions and this gives one natural choice of the
      vector field $X$, namely  $X = \nabla H / \|\nabla H\|^2$.
      In the case of Euclidean space and Lebesgue measure
      this leads to the formula 
      $1/T(E) = \langle \nabla \cdot \frac{\nabla H}{\|\nabla H\|^2}; E\rangle$
      as was found in \cite{Rugh}.
\item Iteration:
      Given the analytic expression for a vector field $X$ and of $\phi$
      as in the Theorem
      the above formulas may be iterated indefinitely.
     Thus formula (\ref{eq:av}) implies that any energy derivative
     of a $\mu$-average can be measured within the $\mu$-ensemble.
     In particular, any derivative
     of the temperature, e.g. specific heat, is also measurable in
     the $\mu$-ensemble.
\item Fluctuations:
      Using the identity  $\mbox{div} (\phi X) = (X \cdot \nabla) \phi
               + \phi \; \mbox{div} X$ we may rewrite the equation
       (\ref{eq:av}) as follows~:
       \begin{equation}
           \frac{\partial}{\partial E} \langle \phi; E\rangle 
          = 
            \langle  (X\cdot \nabla) \phi ; E\rangle
             + \langle  \delta \phi \; \delta (\mbox{div} X) ; E\rangle .
                 \label{eq:avtwo}\end{equation}
      where $\delta \phi = \phi - \langle \phi  ; E\rangle$ etc.
      Thus the energy derivative has a contribution coming from the
      derivative of $\phi$ in the direction of $X$ as well as from
      the product of fluctuations in the observables $\phi$ and
      $\mbox{div} X$, cf.\ example B below.
\end{itemize}
       
\section{Example A : The canonical ensemble}
As an illustration of the Theorem we will show that the canonical
ensemble can be measured through $\mu$-averages.
The canonical ensemble is most conveniently here defined 
as the Laplace transform of the 
$\mu$-ensemble \cite{Bailyn,Becker}. 
More precisely, the weighted partition
function is given by~:
   \begin{equation}
          Z(\beta,\phi) = \int m \; e^{-\beta H} \phi
                      = \int dE \; e^{-\beta E} \mu_E(\phi) .\end{equation}
Assuming that $\mu_E(\phi)$ has an analytic extension
in $E$ and $s$ is small enough we may expand the partition
function by iterating
(\ref{eq:diver}). It is convenient to write
$D_X (\phi) \equiv \mbox{div}\;(\phi X)$ in terms of which
      \begin{equation}
          \mu_{E+s}(\phi) = e^{s \frac{\partial}{\partial E}}
         \mu_E(\phi) = \mu_E (e^{s \; D_X} \phi) .\end{equation}
Here 
the last exponential should be interpreted through 
its power series.
Therefore, formally~:
      \begin{equation}
          Z(\beta,\phi) = \int ds e^{-\beta (E+s)} \mu_E 
             (e^{s \;D_X } \phi) ,\end{equation}
where in this last expression $E$ has a fixed value.
Dividing (normalizing) by the factor $\mu_E(1)=e^{S(E)}$ 
we see that all the terms in the
expansion of the exponential can be measured in the
$\mu$-ensemble. Thus, up to a constant factor
the canonical partition function can in principle be evaluated 
in the $\mu$-ensemble. Of course, this approach does raise the question
about convergence.
In general one can only
hope for the expansion to be asymptotic, e.g. in the number of
degrees of freedom in the system.
To make our statements more explicit
consider the standard partition function ($\phi\equiv 1$)
and the following expansion of the $\mu$-entropy~:
      \begin{equation}
          S(E) = S(E_0) + \sum_{k>0} \frac{(E-E_0)^k}{k!} s_k .
       \label{eq:sk} \end{equation}
As a consequence of Theorem 1 all the numbers $s_k$
(but not $S(E_0)$) are 
measurable in the $\mu$-canonical ensemble.
We shall assume in the following that $s_2$ is strictly negative\footnote
        {This corresponds to the assumption of thermodynamic
         stability, i.e. that the specific
         heat should be positive in the $\mu$-ensemble.
         At least in the thermodynamic limit
         there are mathematical arguments justifying such an assumption.}
The partition function we write as~:
     \begin{equation}
     Z(\beta) = \int dE \; e^{-\beta E + S(E)} = e^{-F(\beta)} .\end{equation}
Now, let $\beta$ be close to $s_1 = \partial S/\partial E (E_0)$ and
let $\Gamma(\beta)$ be the extremal point (the conjugated variable
in the Legendre transform) of the exponent $-\beta E + S(E)$.
Again it is possible to calculate  $\Gamma(\beta)$  perturbatively from
the known quantities $s_k$ and equation (\ref{eq:sk}).
$\Gamma(\beta)$ satisfies (cf. also Bailyn \cite[Section 11.6]{Bailyn})~:
    \begin{equation}
      \beta  = \frac{\partial S}{\partial E} (\Gamma(\beta)) 
                  \label{eq:S} , \ \ \
      1 = \frac{\partial^2 S}{\partial E^2} (\Gamma(\beta))
              \frac{\partial \Gamma}{\partial \beta}  .  \end{equation}
Inserting $E = \Gamma(\beta)+z$ we get for the exponent~:
    \begin{equation}
           -\beta \Gamma(\beta) + S(\Gamma(\beta)) +
          \frac{\partial^2 S}{\partial E^2} (\Gamma(\beta)) z^2 
           \frac{1}{2} + V(z)  \label{eq:V}\end{equation}
where $V(z) = v_3 z^3/3! + v_4 z^4/4! + ...$ can also be calculated in
terms of the coefficients $s_k$.
 As the second derivative was assumed negative,
we may carry out the Gaussian integral by standard
techniques to obtain the (exact) asymptotic formula~:
    \begin{equation}
           F(\beta)  = F_{cl}(\beta)
        -  \log [e^{{\frac{1}{2}}(-\frac{\partial \Gamma}{\partial \beta})
                       (\partial_x)^2} e^{V_\beta(x)}\mbox{}_{|x\equiv 0}] .
           \label{eq:pert}
    \end{equation}
with 
    \begin{equation}
        F_{cl}(\beta) = \mbox{const} +\beta \Gamma(\beta) -  S(\Gamma(\beta))
     - \frac{1}{2} \log (-\frac{\partial \Gamma}{\partial \beta} ).
      \label{eq:F} \end{equation}
For the sake of clarity we shall in the following neglect the $V$-term
and use the last expression (corresponding to the classical action in QFT).
Taking a $\beta$ derivative we get
for the average energy in the canonical ensemble~:
    \begin{equation}
          \langle H; \beta\rangle = F'_{cl}(\beta) = \Gamma(\beta)
        - \frac{1}{2} \frac{\Gamma''(\beta)}{\Gamma'(\beta)}  ,\end{equation}
Differentiating once more we obtain the approximative formula
for the specific heat~:
    \begin{equation}
         c(\beta) = -\beta^2 \frac{\partial 
           \langle H; \beta\rangle}{\partial \beta}
         = -{\beta^2} \Gamma'
               [1 - \frac{\Gamma'''}{2 (\Gamma')^2}
                  + \frac{(\Gamma'')^2}{2 (\Gamma')^3} ] .\end{equation}
As a trivial but analytically accessible
example (cf. \cite{Rugh})
 we consider $N$ harmonic oscillators for which the
$\mu$-entropy equals $S(E) = (N-1) \log E$
and the extremum  of $-\beta E + S(E)$ 
is attained for $\Gamma(\beta)= (N-1)/\beta$.
We get~:
      \begin{equation}
          \langle H; \beta\rangle = \frac{N-1}{\beta} -
             \frac{1}{2} \frac{ {- 2 (N-1)}/{\beta^3}}
                 {(N-1)/{\beta^2}} =  \frac{N}{\beta} ,\
           c(\beta) = N ,\end{equation}
which happens to recover the exact canonical results.\\

\section{Example B : Interacting particles }
Our second application is concerned with the thermodynamics of
$N$ particles in a box  of volume $V$ in  $R^3$. 
We shall compare the micro-canonical computations, e.g.\ of
temperature,
given here with formulae obtained from the canonical ensemble theory.
It turns out that a natural assumption on the fluctuations in the kinetic
energy is sufficient to obtain equivalence of the two approaches.

The Hamiltonian is taken to be of the form
\begin{equation}
  H(p,q) = \sum_{i=1}^{3N} p_i^2/2 + U_{\rm int}(q) + U_{\rm ext}(q),
\end{equation}
with the standard symplectic structure on $R^{6N}$.
Here $K(p) = \sum_i p_i^2/2$ is the kinetic energy,
$U_{\rm ext}=\sum_{j=1}^N U_j(\vec{q}_j)$ is a box confining potential
and $U_{\rm int}$ is an interaction potential, e.g.\
a sum of  two-body interactions.
Note that for notational convenience the momentum vector is here considered
as a $3N$ dimensional vector whereas the configuration coordinates
are considered as $N$ three-dimensional vectors.\\

For the vector field $X$ we choose~:
$X_1 = \vec{p}/2K(p)$. The reader might worry about the fact that 
this vector field is singular at $\vec{p}=\vec{0}$\ but as we
shall see below the singularity is integrable when
the number of particles is sufficiently large (at least 2 particles
are needed in the applications below).
A direct computation shows that~:

\begin{equation}
 \mbox{div}\; (X_1) = (3N -2)/ (2 K(p)) .
\end{equation}
The temperature is then given by~:
\begin{equation}
   \frac{1}{T(E)} = \frac{3N-2}{2} \langle 1/K(p) ; E \rangle.
\end{equation}
In order for the average to be well-defined it is necessary that
$1/K(p)$ is integrable at $p=0$, i.e.\ that
$\int 1/p^2 d^{3N}p < \infty$ where the integral is over a 
neighborhood of $p=0$. This is the case when
$3N > 2$,
i.e.\ when the system contains at least one particle.

The inverse specific heat, $1/c = \frac{\partial}{\partial E}
   T(E) = - T(E)^2 \frac{\partial}{\partial E} \frac{1}{T(E)}$,
can be calculated using (\ref{eq:av}) from which
\begin{equation}
  ( \frac{1}{T(E)} + \frac{\partial}{\partial E}) \; \frac{1}{T(E)}
    = \langle \mbox{div} (\frac{3N-2}{K} X_1 );E \rangle
    =  \frac{(3N-2)(3N-4)}{4}\langle{1/K^2} ;E\rangle
\end{equation}
and thus
\begin{equation}
   1/c(E) = 1 - \frac{(3N-4) \langle 1/K^2 ; E \rangle}
             {(3N-2) \langle 1/K ; E \rangle^2} .
   \label{eq:specheat}
\end{equation}
This time we need that $1/K^2(p)$ is integrable and this happens
when $3N>4$,
 i.e.\ at least two particles are present.

The standard definition of the (canonical) temperature is
    \begin{equation}
         T_c(E) = \frac{2}{3N} \langle K(p); E \rangle .
    \end{equation}
Had the average been in the canonical ensemble this would indeed have
been the canonical temperature.
We have the following formula for the ratio~:
    \begin{equation}
          T_c(E)/T(E) = \frac{3N-2}{3N} \langle K;E \rangle
                          \langle 1/K;E \rangle .
          \label{eq:t}
    \end{equation}
We may also calculate the inverse specific heat this time using the
canonical temperature.  From (\ref{eq:av}) and the obvious
 identity $\mbox{div} (K X_1) \equiv 3N/2$
we obtain~:
    \begin{equation}
         \frac{1}{c_c(E)} = \frac{\partial}{\partial E} T_c(E) = 
                  1 - \frac{T_c(E)}{T(E)} .
          \label{eq:tc}
    \end{equation}

In order to compare the above formulas we shall consider
the fluctuations in the kinetic energy, defined by
   \begin{equation}
        K = \langle K(p);E \rangle + \delta K.
   \end{equation}
Our assumption in the following will be that
$\langle K \rangle = \langle K(p);E \rangle$
and $(\delta K)^2$ are both of order $N$. In particular, that for
large enough $N$  the singularities are integrable and we have
the expansion~:
   \begin{equation}
         \frac{1}{K} = \frac{1}{\langle K \rangle}
                     - \frac{\delta K}{\langle K \rangle^2}
                     + \frac{(\delta K)^2}{\langle K \rangle^3}
                     + o(N^{-2}).
   \end{equation}
Taking the average on both sides yields~:
   \begin{equation}
        \langle \frac{1}{K} \rangle =
              \frac{1}{\langle K \rangle} +
              \frac{\langle (\delta K)^2 \rangle}{\langle K \rangle^3}  
              + o(N^{-2}) 
    \label{eq:averone}
   \end{equation}
and similarly by squaring before taking the average~:
   \begin{equation}
        \langle \frac{1}{K^2} \rangle =
              \frac{1}{\langle K \rangle^2} +
              3 \frac{\langle (\delta K)^2 \rangle}{\langle K \rangle^4}  
              + o(N^{-3}) 
    \label{eq:avertwo}
    \end{equation}
Using (\ref{eq:averone})-(\ref{eq:avertwo})
 and retaining only terms to order $N^{-1}$
both (\ref{eq:specheat}) and (\ref{eq:tc}) reduces to
    \begin{equation}
         \frac{1}{c(E)} = \frac{2}{3N}
               - \frac{\langle (\delta K)^2 ; E \rangle}
                      {\langle K ; E            \rangle}
               + o(N^{-1}) .
    \end{equation}
We have here given a micro-canonical derivation of an
expression which relates the specific heat to fluctuations
in the kinetic energy. This was previously found by Lebowitz
{\it et al} \cite{LPV} using an ingenious technique of inverting
ensemble averages. Their method, however, relies on (unstated)
analytic properties of the ensembles involved whereas the method
presented here makes the assumptions explicit in terms
of the fluctuations in the kinetic energy.
We also note that our assumption on $\delta K$ gives a sufficient
condition  for the equivalence not only of $T(E)$ and $T_c(E)$ (the
 ratio (\ref{eq:t}) differs from 1 by a term of
order $N^{-1}$) but also for the derived expressions for
the inverse specific heat (to order $N^{-1}$).
Even close to a phase transition where $1/c(E)$
tends to zero we would expect that our assumptions
(\ref{eq:averone})-(\ref{eq:avertwo}) are not violated.\\

As a final application consider the $\mu$-canonical pressure
exerted by the $N$ particles on the walls of the container.
We may define it as the average force per surface area of the
container. The pressure, $P$, can then be calculated through the
Virial Theorem
(see e.g. Becker \cite[p. 98]{Becker} or Abraham and Marsden
 \cite[Example 3.7.32]{Abraham}).
The $j$'th particle $j=1,\ldots,N$ having coordinates $q_j$
is confined by the external potential 
$U_j(q_j)$.
 The force, $d\vec{F}= P \; d\vec{A}$, exerted
on the  surface element $d\vec{A}$ is given
by the (time-) average of the external forces in a small neighborhood
$\delta V$ of that surface element, i.e.
$P \; d\vec{A} = \langle \sum_j \vec{\nabla} U_j : q_j \in \delta V \rangle$.

This formula is vectorial but a scalar quantity is obtained
by taking  the scalar product
with the coordinate of the volume element in question.
Summing over the whole surface removes the restriction on the
$q_j$'th coordinate, thus yielding 
 \begin{equation}
      P \int \vec{q} \cdot d\vec{A} =  \langle 
   \sum_j \vec{q}_j \cdot \vec{\nabla} U_j  \rangle 
   \end{equation}
and finally, using Stokes Theorem we see that the left hand side equals $3PV$.
By the Virial Theorem the time average of
$(p \partial_p - q \partial_q)H$ vanishes (since it is a total
time derivative). Hence
     \begin{equation}
         PV/N = \frac{1}{3N} \langle 2 K(p) - \Phi; E \rangle,
     \end{equation}
with $\Phi = \sum_{j=1}^N \vec{q}_j \cdot {\partial U_{\rm int}(q)}
/{\partial \vec{q}_j} $.   
    
We therefore obtain the following derivative~:
     \begin{equation}
          \frac{\partial}{\partial T} (PV/N)
         =  c(E) \frac{\partial}{\partial E} (PV/N) 
         = 1 - \frac{c(E)}{3N} \frac{\partial}{\partial E} 
                \langle \Phi ; E \rangle .
     \end{equation}
Equation (\ref{eq:avtwo}) implies the exact formula~:
   \begin{equation}
         \frac{\partial}{\partial T} (PV/N-T) =
            c(E) \frac{3N-2}{6N} \langle \delta \Phi \; \delta (1/K) 
             ; E \rangle , 
   \end{equation}
which under the assumption above on the kinetic energy fluctuations
reduces to
   \begin{equation}
         \frac{\partial}{\partial T} (PV/N-T) =
            \frac{2 c(E)}{(3NT(E))^2} \langle \delta \Phi \; \delta K 
             ; E \rangle + o(1) . 
   \end{equation}
This again is a micro-canonical 
rederivation of a formula previously obtained
by Lebowitz {\it et al} \cite{LPV} under the 
afore-mentioned analyticity assumption
of ensemble inversion.\\
           
I am grateful to Cristian Giardina, John Milnor, Ettore Aldrovandi and
Dietmar Salamon among others
who have helped me clarifying the ideas and
calculations presented here.

\end{document}